\documentclass[prd,aps,showpacs,nofootinbib]{revtex4} 
\begin{document}
\def \a {\alpha}
\def \b {\beta}
\def \g {\gamma}
\def \G {\Gamma}
\def \d {\delta}
\def \eps {\varepsilon}
\def \ep {\epsilon}
\def \e {\eta}
\def \f {\phi}
\def \ffi {\varphi}
\def \j {\iota}
\def \th {\theta}
\def \vth {\vartheta}
\def \k {\kappa}
\def \l {\lambda}
\def \m {\mu}
\def \n {\nu}
\def \x {\xi}
\def \p {\pi}
\def \r {\rho}
\def \s {\sigma}
\def \t {\tau}
\def \ps {\psi}
\def \o {\omega}
\def \z {\zeta}
\def \L {\pounds}
\def \vu {\tilde u}
\def \der {\partial }
\def \nn {\nonumber}
\def \rov {\equiv}
\def \A {{\cal A}}
\def \BB {{\cal B}}
\def \C {{\cal C}}
\def \D {{\cal D}}
\def \E {{\cal E}}
\def \F {{\cal F}}
\def \GG {{\cal G}}
\def \K {{\cal K}}
\def \N {{\cal N}}
\def \L {{\cal L}}
\def \cN {\bar\N}
\def \vN {\tilde\N}
\def \M {{\cal M}}
\def \vM {\tilde \M}
\def \I {{\cal I}}
\def \J {{\cal J}}
\def \R {{\cal R}}
\def \CC {{\rm C}}
\def \U  {{\cal U}}
\def \T  {{\cal T}}
\def \bfi {\bar \varphi}
\def \br {\bar \rho}
\def \bz {\bar z}
\def \bt {\bar t}
\def \pul {{{\scriptstyle{\frac{1}{2}}}}}
%----------------------------------------------trig fce
\def \sn {\sin \th}
\def \cs {\cos \th}
\def \tg {\tan \th}
\def \ctg {\cot \th}
\def \csec {\csc \th}
\def \dcsec {\csc^2 \th}
\def \msn {\sin^{-1} \th}
\def \ct {\bar t}
\def \dsn {\sin^2 \th}
\def \mdsn {\sin^{-2} \th}
\def \tsn {\sin^3 \th}
\def \csn {\sin^4 \th}
\def \dcs {\cos^2 \th}
\def \tcs {\cos^3 \th}
\def \csin {\sqrt{1-(wu)^2}}
\def \dctg {\cot^2 \th}
\def \chd  {\cosh 2\d}
\def  \shd  {\sinh 2\d}
%----------------------------------------mezery
\def \mm {\mbox{\quad }}
\def \mv {\mbox{\qquad }}
\def \msip {\rightarrow}
\def \vsip {\longrightarrow}
\def \lkz  {\bigl(}
\def \pkz  {\bigr)}
\def \lvkz {\Bigl(}
\def \pvkz {\Bigr)}
\def \lvvkz {\biggl(}
\def \pvvkz {\biggr)}
\def \lhz  {\bigl[}
\def \phz  {\bigr]}
\def \lvhz {\Bigl[}
\def \pvhz {\Bigr]}
\def \lvvhz {\biggl[}
\def \pvvhz {\biggr]}
\def \lsz   {\bigl\{ }
\def \psz   {\bigr\} }
\def \pvsz {\Bigl\} }
\def \lvsz {\Bigr\{ }
\def \lvvsz {\Biggl\{}
\def \pvvsz {\Biggr\}}
%-----------------------------------------------
%--------------------------------------matematicke  prostredi
\def \BE {\begin{equation}}
\def \EE {\end{equation}}
\def \BDM {\begin{displaymath}}
\def \EDM {\end{displaymath}}
\def \BEAH {\begin{eqnarray*}}
\def \EEAH {\end{eqnarray*}}
\def \BEA {\begin{eqnarray}}
\def \EEA {\end{eqnarray}}
\def \BM {\begin{math}}
\def \EM {\end{math}}
\def \BDM {\begin{displaymath}}
\def \EDM {\end{displaymath}}
%----------------------------------------mezery
\def \mm {\mbox{\quad }}
\def \mv {\mbox{\qquad }}
\def \msip {\rightarrow}
\def \vsip {\longrightarrow}
%--------------------------------------------elmg doln¡
\def \Fab {F_{\a \b }}
\def \Fmn {F_{\m \n }}
\def \Fgb {F_{\g \b }}
\def \Fag {F_{\a \g }}
\def \Fnn {F_{00}}
\def \Fnj {F_{01}}
\def \Fnd {F_{02}}
\def \Fnt {F_{03}}
\def \Fjn {F_{10}}
\def \Fjj {F_{11}}
\def \Fjd {F_{12}}
\def \Fjt {F_{13}}
\def \Fdn {F_{20}}
\def \Fdj {F_{21}}
\def \Fdd {F_{22}}
\def \Fdt {F_{23}}
\def \Ftn {F_{30}}
\def \Ftj {F_{31}}
\def \Ftd {F_{32}}
\def \Ftt {F_{33}}
%---------------------
%-----------------------------------------------
\newcommand{\zl}[2]{{{\scriptstyle{\frac{#1}{#2}}}}}
\def \pul {{{\scriptstyle{\frac{1}{2}}}}}
\def \tripul {{{\scriptstyle{\frac{3}{2}}}}}
\def \ctvrt {{{\scriptstyle{\frac{1}{4}}}}}
\def \osmina {{{\scriptstyle{\frac{1}{8}}}}}
\def \sestina {{{\scriptstyle{\frac{1}{6}}}}}
%-----------------------------------
\def \VUW {\left(\V\edb-r^2\edg U^2\chd
           -r^2\emdg W^2\chd-2r^2 UW\shd\right)}
\def \V  {\frac{V}{r}}
\def \edb  {{\rm e}^{2\b }}
\def \edg  {{\rm e}^{2\g }}
\def \emdg {{\rm e}^{-2\g }}
\def \cc {(c,_\th+2c\ctg)}
\def \dd {(d,_\th+2d\ctg)}
%-------------------------------------------metrika dolni
\def \gab {g_{\a \b }}
\def \gmn {g_{\m \n }}

\title{
Boost-rotation symmetric 
vacuum spacetimes with spinning sources}
\author{A. Pravdov\' a}
\email{pravdova@math.cas.cz}
\author{V. Pravda}
\email{pravda@math.cas.cz}
\affiliation{Mathematical Institute, 
Academy of Sciences, 
\v Zitn\' a 25,
115 67 Prague 1, Czech Republic }

\begin{abstract}
Boost-rotation symmetric vacuum spacetimes with spinning sources
which correspond
to gravitational field of uniformly accelerated spinning ``particles''
are studied. Regularity conditions and asymptotic properties
are analyzed. 
News functions are derived by transforming the~general 
spinning boost-rotation
symmetric vacuum metric to Bondi-Sachs coordinates. 
\end{abstract}
\maketitle

%--------------------------------------------

\section{Introduction and Summary}

Boost-rotation symmetric spacetimes correspond to gravitational field
of uniformly accelerated ``particles''. Usually conical singularities,
which provide the~``source'' of the~acceleration, 
are also present on the~axis of the~axial symmetry.

Boost-rotation symmetric  spacetimes have two Killing vectors 
(the~axial $\x$ and the~boost $\e$ Killing vectors) 
and it has been proven that they are the~only 
axially symmetric spacetimes with an additional symmetry that are radiative 
and admit global null infinity \cite{jibisch}.
This result was generalized for spinning sources, i.e.
for non-hypersurface orthogonal
 Killing vectors    in \cite{ajajibi}.
Moreover boost-rotation symmetric spacetimes are the only radiative asymptotically flat
spacetimes known in an analytical form which represent moving sources.
While there are several known boost-rotation symmetric solutions 
with non-rotating sources
(see \cite{AV,JibiEhlers} and references therein), there is
only one known exact solution with spinning sources -- the~spinning
C-metric \cite{PlebDem,bivoj,letol} corresponding to two uniformly accelerated
Kerr black holes.

Thanks to the~rotation of sources there appear torsion singularities 
besides conical singularities  on the
axis of the~axial symmetry and consequently there can be regions with closed timelike
curves (see \cite{let} and also \cite{bivoj,bonnor} for examples and \cite{bonnor} for 
references). 

The~structure of a boost-rotation symmetric spacetime with hypersurface 
orthogonal Killing vectors was studied in \cite{BicSchPRD}
and the~general form  of its news function was found 
in \cite{bicakBS,BicTN}.
Recently, news functions for spinning boost-rotation symmetric
Petrov type D spacetimes were computed in late time 
approximation in \cite{kroon}.

The~present paper, where some results by {Bi\v c\'ak} and
{Bi\v c\'ak} \& Schmidt from \cite{BicSchPRD,bicakBS,BicTN}
 are generalized, is organized as follows.
In Sec.~\ref{secasympreg} 
spinning boost-rotation symmetric (brs)
vacuum spacetimes in 
coordinates adapted to the~boost and rotation symmetries and null 
coordinates
are examined, e.g. 
regularity of the~spacetime 
on the~roof and on the~axis and asymptotic flatness at null infinity
are studied.
In Sec.~\ref{secbondi} the~spinning brs metric is transformed
from the~coordinates
adapted to the~boost
and rotation symmetries 
to the~Bondi-Sachs coordinates 
\cite{bondi,sachs,burg}, suitable for examining radiation,
 to find the~news functions
of spinning brs spacetimes.

%-------------------------------------

%-----------------------------------------

\section{Spinning boost-rotation symmetric spacetimes --
regularity conditions, asymptotic behaviour}
\label{secasympreg}

%-----------------------------
The general form of spinning brs metric 
in coordinates adapted to the~boost and rotation symmetries
$\{ t$, $\r$, $z$, $\ffi\}$ is  \cite{bivoj}
\BEA
{\rm d}s^2&=&-{\rm e}^\l {\rm d}\r^2 -\r^2{\rm e}^{-\m}{\rm d}\ffi^2 
+2a{\rm e}^\m (z{\rm d}t-t{\rm d}z){\rm d}\ffi
+a^2{\rm e}^\m(z^2-t^2){\rm d}\ffi^2\ \nn\\
&&
-\frac{1}{z^2-t^2}[({\rm e}^\l z^2-{\rm e}^\m t^2) {\rm d}z^2
                  -2zt ({\rm e}^\l-{\rm e}^\m) {\rm d}z\ {\rm d}t
                  -({\rm e}^\m z^2-{\rm e}^\l t^2){\rm d} t^2]
\ ,\label{dsbrs}
\EEA
where 
$\m$, $\l$, and $a$ are functions of 
\BDM
A=\r^2, \mm B=z^2-t^2\ .\nn 
\EDM
It has two Killing vectors
\BE
\x=\frac{\der}{\der\ffi} \ ,\mm 
\e=t\frac{\der}{\der z} +z\frac{\der}{\der t} \label{Killv}
\EE
with norms
\BEA
      \x^\a\x_\a&=&g_{\ffi\ffi}=-\r^2{\rm e}^{-\m}+a^2(z^2-t^2){\rm e}^\m
                 =-A {\rm e}^{-\m}+a^2B{\rm e}^\m\ ,\nn\\
      \e^\a\e_\a&=&g_{tt}z^2+g_{zz}t^2+2g_{zt}zt=B{\rm e}^\m\ .\nn
\EEA
As in the~non-spinning case \cite{BicSchPRD}, 
two null hyperplanes $B=0$, i.e. $z=\pm t$, will be 
called  the~``roof'', the~points with $A=0$ the~``axis'',
the~region of the~spacetime with $B<0$ ``above the~roof'', 
and finally the~region with $B>0$ ``bellow the~roof''.
Notice that the~behaviour of the~boost and axial Killing vectors
(\ref{Killv})
is more complicated in the~spinning case. Bellow the~roof
($B>0$), the~boost Killing vector $\e$ 
is mostly timelike  as in the~non-spinning case
but in the~vicinity of spinning sources there may also occur 
ergoregions where it is spacelike.
Due to the~presence of spinning strings 
there may be also regions in their neighbourhood
with closed timelike curves where
the~axial Killing vector $\x$ is timelike. 
In order to determine if there exist both timelike and spacelike 
Killing vectors everywhere bellow the~roof ($B>0$),
 we study a general linear combination of the~boost
and the~axial Killing vectors with constant coefficients 
$X=c_1\x+c_2\e$.
Its norm 
may be both positive and negative if the~product of
eigenvalues of the~quadratic form 
(${c_1}^2g_{\f\f}+\dots$) given by the~norm is negative, i.e. if
$-\r^2 B<0$,
which is satisfied everywhere bellow the~roof, where the~spacetime
is thus stationary and may
be transformed to the~stationary Weyl metric 
(\ref{dsWeyl}) (see e.g. \cite{bivoj}).
However, above the~roof ($B<0$), 
the~product is everywhere positive 
$-\r^2 B>0$ and thus there
does not exist a timelike Killing vector and the~spacetime
is nonstationary there.\footnote{It is easier to perform these
calculations in coordinates $\{ \g,\ \r,\ \b,\ \ffi\}$ 
and $\{ b,\ \r,\ \chi,\ \ffi\}$, given in App.~\ref{ap-metric}, for regions
bellow and above the~roof, respectively.}

%--------------------------------

Vacuum Einstein's equations for the~spinning brs metric (\ref{dsbrs}) are 
\BEA
A\m,_{AA}+B\m,_{BB}+\m,_A +\m,_B
&=&-\frac{B}{A}\ {\rm e}^{2\m}  (A{a,_A}^2+B{a,_B}^2)\ ,\label{Ercemu}\\
0&=&AB\lvkz {\rm e}^{2\m}a,_A\pvkz,_A
   +\lvkz B^2{\rm e}^{2\m}a,_B\pvkz,_B\ ,\label{Ercea}\\
(A+B) \l,_A &=& (A-B)\m,_A-2B\m,_B-B(B{\m,_B}^2-A{\m,_A}^2)
   +2AB\m,_A\m,_B
\nn\\  &&
           +\frac{B^2}{A}{\rm e}^{2\m}(B{a,_B}^2-A{a,_A}^2-2A a,_A a,_B)
\ ,   \label{ErcelA}\\
(A+B) \l,_B &=& (A-B)\m,_B+2A\m,_A+A(B{\m,_B}^2-A{\m,_A}^2)    
+2AB\m,_A\m,_B
\nn\\&&
           -B{\rm e}^{2\m}(B{a,_B}^2-A{a,_A}^2+2B a,_A a,_B)\ .
\label{ErcelB}
\EEA
Notice that Eqs.~(\ref{Ercemu}), (\ref{Ercea}) are integrability 
conditions for Eqs.~(\ref{ErcelA}), (\ref{ErcelB}).

%----------------------------

First let us investigate the~regularity of the~roof and the~axis.
From Eq.~(\ref{ErcelA}) it follows that on the~roof (i.e. for $B=0$)
\BDM
\l,_A(A,0) =\m,_A(A,0) \mm \msip\mm \l (A,0) -\m(A,0)=K_1=\mbox{const}
\ .\nn
\EDM

The~roof is regular (i.e. $g_{zz}$, $g_{tt}$, and $g_{tz}$ in (\ref{dsbrs})
are nonsingular on the~roof) if for $B=0$
\BE
\l(A,0) =\m(A,0)\ , \mm
% ,\mm \der^{(n)}_A \l(A,0) =\der^{(n)}_A\m (A,0) \ , \mm
\mbox{i.e.}\mm K_1=0\ .\label{regroof}
\EE

%-----------------------------------------------

From Eqs.~(\ref{Ercemu}), (\ref{ErcelA}), and (\ref{ErcelB})  
on the~axis ($A=0$) we get
\BEA
a,_B(0,B) &=&0\mm \msip\mm a={\tilde a}_{0} +{\tilde a}_1(B) A
+{\cal O}(A^2)\ ,\mm {\tilde a}_0=\mbox{const}\ ,\nn\\
\l,_B(0,B) +\m,_B(0,B) &=&0\mm \msip \mm 
\l(0,B) +\m(0,B) =K_2=\mbox{const}\nn
\ .
\EEA

The~axis regularity condition
\BDM
{\lim_{\r_0\msip0} \frac{1}{2\p} 
        \frac{\int_0^{2\p} \sqrt {g_{\ffi\ffi}|_{\r_0} } {\rm d}\ffi}
             {\int_0^{\r_0}\sqrt{g_{\r\r}}{\rm d}\r}=1} \nn  \
\EDM
(or equivalently $g_{xx}$, $g_{yy}$ are nonsingular and  $g_{xy}=0$ there, 
see (\ref{dsxy}) in App.~\ref{ap-metric})
is satisfied if
\BEA
a(0,B) &=&0\ ,\mm %\der^{(n)}_B a(0,B) =0\mm
\msip\mm a={\tilde a}_1 (B) A+{\cal O}(A^2)\ ,
\mm \mbox{i.e.}\mm {\tilde a}_{0}=0\ , \label{regosa-a}\\
\l(0,B) +\m(0,B) &=&0\ ,
\mm \mbox{i.e.}\mm  K_2=0\ .\label{regosaml}
%\label{regosakonst}
\EEA
If $K_2\not= 0$ there is a conical singularity 
along the~axis and if ${\tilde a}_{0}\not= 0$ a torsion singularity
is present there. The~regularity condition of the~roof (\ref{regroof})
is the~same as for nonspinning brs spacetimes \cite{BicSchPRD}, however,
a new condition (\ref{regosa-a}) arises for the~regularity of the~axis
except for (\ref{regosaml}) which also appears in the~nonspinning case
\cite{BicSchPRD}.

Now let us turn our attention to 
asymptotic behaviour of spinning brs spacetimes at null infinity.
For this purpose we transform 
(\ref{dsbrs}) to null coordinates 
 in two steps: first  transforming it 
to coordinates $\{ b$, $\r$, $\chi$, $\ffi\}$ by (3.10) in \cite{BicSchPRD}
\BDM
b=\sqrt{-B}=\sqrt{t^2-z^2}\ ,\mm \tanh\chi =\pm \frac{z}{t}\ \nn
\EDM
we obtain the~metric 
\BE
{\rm d}s^2={\rm e}^\l ({\rm d}b^2 -{\rm d}\r^2)
-\r^2{\rm e}^{-\m}{\rm d}\ffi^2 
-b^2{\rm e}^{\m}({\rm d}\chi+a{\rm d}\ffi)^2\ .\label{dsbchi}
\EE
Finally transforming (\ref{dsbchi}) 
to coordinates $\{ {\bar u}$, ${\bar v}$, $\chi$, $\ffi\}$  by (3.15)
in \cite{BicSchPRD}
\BDM
{\bar u}=b-\r\ ,\mm {\bar v}=b+\r\ \nn \
%,\mm \mbox{or inverse} \mm \r=\pul (v-u)\ ,\mm b=\pul (v+u) 
\EDM
we obtain %the~metric has the~form
\BE
{\rm d}s^2={\rm e}^\l {\rm d}{\bar u}{\rm d}{\bar v} 
-\frac{({\bar v}-{\bar u})^2}{4}{\rm e}^{-\m}{\rm d}\ffi^2 
-\frac{({\bar v}+{\bar u})^2}{4}{\rm e}^{\m}({\rm d}\chi+a{\rm d}\ffi)^2
\ .\label{dsuvchi}
\EE
Vacuum Einstein's equations for (\ref{dsuvchi}) read
\BEA
\m,_{{\bar u}{\bar v}}+\frac{1}{{\bar v}^2-{\bar u}^2}
({\bar v}\m,_{\bar u}-{\bar u}\m,_{\bar v})
&=&\lvkz \frac{ {\bar v}+{\bar u}}{{\bar v}-{\bar u}}\pvkz^2 {\rm e}^{2\m} 
 a,_{\bar u} a,_{\bar v}\ ,\nn\\
0&=&a,_{{\bar u}{\bar v}}+a,_{\bar u} \lvkz \m,_{\bar v}
+\frac{{\bar v}-2{\bar u}}{{\bar v}^2-{\bar u}^2}\pvkz
          +a,_{\bar v}  \lvkz \m,_{\bar u}
       +\frac{2{\bar v}-{\bar u}}{{\bar v}^2-{\bar u}^2}\pvkz\ ,\nn\\
-{\bar u} \l,_{\bar u} &=& {\bar v}\m,_{\bar u}
      +\frac{{\bar v}^2-{\bar u}^2}{4}{\m,_{\bar u}}^2
            +\frac{({\bar v}+{\bar u})^3}{{\bar v}-{\bar u}}
            \frac{{\rm e}^{2\m}}{4}{a,_{\bar u}}^2\ ,\label{Erceuvchi}
\\
-{\bar v} \l,_{\bar v} &=&{\bar u}\m,_{\bar v}
       -\frac{{\bar v}^2-{\bar u}^2}{4}{\m,_{\bar v}}^2
            -\frac{({\bar v}+{\bar u})^3}{{\bar v}-{\bar u}}
            \frac{{\rm e}^{2\m}}{4}{a,_{\bar v}}^2\ .\nn
\EEA

%-----------------------------------------

Assuming the~metric functions $\m$, $\l$, and $a$ to have expansions in ${\bar v}^{-1}$
for ${\bar v}\msip\infty$ ($\m({\bar u},\ {\bar v})
=\m_0({\bar u})+\m_1({\bar u})/{\bar v}+\dots $) and solving 
Eqs.~(\ref{Erceuvchi})  at null infinity, i.e. 
for the~limit ${\bar v}\msip\infty$  and ${\bar u}$, $\chi$,
$\ffi$ constant, we get 
\BEA
\m&=& \m_0+\frac{\m_1({\bar u})}{{\bar v}}+{\cal O}({\bar v}^{-2})\ ,
\nn\\
\l&=&\l_0({\bar u})+\frac{\l_1({\bar u})}{{\bar v}}+{\cal O}({\bar v}^{-2})\ ,
\label{limmla}\\
a&=& a_0+\frac{a_1({\bar u})}{{\bar v}}+{\cal O}({\bar v}^{-2})\ ,
\nn 
\EEA 
where $a_0$ and $\m_0$ are constants and $\l_0(\bar u)$ satisfies
\BDM
\l_0,_{\bar u}=-\frac{1}{4 {\bar u}}\lvkz 4\m_1,_{\bar u}
+{\m_1,_{\bar u}}^2+{\rm e}^{2\m_0}{a_1,_{\bar u}}^2\pvkz\ .\nn
\EDM
The~metric (\ref{dsuvchi}) with the~metric functions (\ref{limmla}) 
is asymptotically Minkowskian at null infinity
as in the~limit ${\bar v}\msip\infty$  and ${\bar u}$, $\chi$,
$\ffi$ constant, it can be
transformed to the~Minkowski metric using the~transformations (3.23), (3.24)
in \cite{BicSchPRD}
\BDM
{\bar u}'={\rm e}^{\pul\m_0}\int {\rm e}^{\l_0({\bar  u})}{\rm d}{\bar u}\ ,
\mm {\bar v}'={\rm e}^{-\pul\m_0}{\bar v}\ ,
\mm  \chi '={\rm e}^{\m_0}\chi\nn
\EDM
and 
\BDM
\chi ''=\chi '+a_0{\rm e}^{\m_0}\ffi\ .\nn
\EDM

%-------------------------------------------

%----------------------------------------------

\section{The~Bondi-Sachs coordinates and news functions 
for spinning brs spacetimes}
\label{secbondi}

In this section we transform the~spinning brs metric (\ref{dsbrs})
into the~Bondi-Sachs coordinates \mbox{$\{ u$,~$r$,~$\th$,~$\f \}$},
in which the~metric, that does not depend on $\f$ 
because of the~axial symmetry, has 
the~form \cite{bondi,sachs,burg}
\BEA
{\rm d}s^2\!\!  &=&\!\!  
g_{uu}{\rm d}u^2+2g_{ur}{\rm d}u{\rm d}r+2g_{u\th}{\rm d}u{\rm d}\th
+2g_{u\f}{\rm d}u{\rm d}\f+g_{\th\th}{\rm d}\th^2+g_{\f\f}{\rm d}\f^2
+2g_{\th\f}{\rm d}\th{\rm d}\f\ \label{ds}
\EEA
with the~following expansion for $r\msip \infty$ and $u$, $\th$, and $\f$ constant
\BEA
g_{uu}&=&1-\frac{2M}{r}+{\cal O}(r^{-2})\ ,\nn\\
g_{ur}&=&1-\frac{c^2+d^2}{2r^2}+{\cal O}(r^{-4})\ ,\nn\\
g_{u\th}&=&-\cc+{\cal O}(r^{-1})\ ,\nn\\
g_{u\f}&=&-\dd \sn+{\cal O}(r^{-1})\ ,\label{expanse}\\
g_{\th\th}&=&-r^2-2cr-2(c^2+d^2)+{\cal O}(r^{-1})\ ,\nn\\
g_{\th\f}&=&-2dr\sn +{\cal O}(r^{0})\ ,\nn\\
g_{\f\f}&=&-r^2\dsn+2cr\dsn-2(c^2+d^2)\dsn+{\cal O}(r^{-1})
\ ,\nn
\EEA
where $c$, $d$, $M$ are functions of $u$ and $\th$.
As a consequence of Einstein's equations, time dependence of the~mass aspect $M$
is determined by the~news functions $c,_u$ and $d,_u$ \cite{sachs,burg}
\BDM
M,_u  =-({c,_u}^2+{d,_u}^2)+\pul(c,_{\th\th}+3c,_\th\ctg-2c),_u\ .\nn 
\EDM
If there is nonvanishing news function, gravitational radiation is present
and the~total Bondi mass at future null infinity is decreasing.

%------------------------------------------

In order to find the~transformation of
spinning brs spacetimes from the~coordinates
$\{ t$, $\rho$, $z$, $\ffi\}$ with the~metric (\ref{dsbrs})
into the~Bondi-Sachs coordinates \mbox{$\{ u$,~$r$,~$\th$,~$\f \}$ }
with the~metric (\ref{ds}) and its  expansions (\ref{expanse})
we follow \cite{bicakBS} and we first transform the~metric (\ref{dsbrs}) to flat-space  
spherical coordinates
$\{ R$, $\Theta$, $\ffi\}$ and a flat-space retarded time $U$ using the~relations
\BEA
t    &=& U+R\ ,\nn\\
\rho &=& R \sin\Theta\ ,\label{trURTh}\\ 
z    &=& R \cos\Theta\ .\nn
\EEA
We assume the~metric functions to have the~expansions 
in powers $R^{-1}$
\BEA
\lambda(U,\ \Theta)&=& \lambda_0 (U,\ \Theta)+ 
\frac{\lambda_1 (U,\ \Theta)}{R}
+{\cal O}(R^{-2})\ ,\nn\\
\m (U,\ \Theta)    &=& \m_0       + \frac{\m_1 (U,\ \Theta)}{R}
+{\cal O}(R^{-2})\ ,\label{rozvlma}\\
a(U,\ \Theta)      &=& a_0 + \frac{a_1 (U,\ \Theta)}{R }
+{\cal O}(R^{-2})\ ,\nn
\EEA
where $\m_0$ and $a_0$ are constants and thus
\BEA
{\rm e}^{\lambda}&=&\beta (U,\ \Theta) \lvkz
       1+\frac{\lambda_1 (U,\ \Theta)}{R}
+{\cal O}(R^{-2}) \pvkz\ ,\nn\\
{\rm e}^{\mu}    &=&\alpha  \lvkz 1+\frac{\mu_1 (U,\ \Theta)}{R}
+{\cal O}(R^{-2}) \pvkz \ \label{rozvelm}
\EEA
with
\BEA
\beta (U,\ \Theta) &=& {\rm e}^{\lambda_0 (U,\ \Theta)}\ ,\nn\\
\alpha             &=&  {\rm e}^{\mu_0}\ .\label{betaalfa}
\EEA

Now we transform the~metric further to the~Bondi-Sachs coordinates 
by an asymptotic transformation
\BEA
U   &=& %\stackrel{ln}{\p}(u,\th)\ln r+
         \stackrel{o}{\p}(u,\th)
      +\frac{\stackrel{1}{\p}(u,\th)}{r}
          +\frac{\stackrel{2}{\p}(u,\th)}{r^2}
         +{\cal O}(r^{-3})   \ ,\nn\\
R   &=&q(u,\th)r+\stackrel{o}{\s}(u,\th)
           +\frac{\stackrel{1}{\s}(u,\th)}{r}+{\cal O}(r^{-2})    \ ,\nn\\
%\vth
\Theta &=&\stackrel{o}{\t}(u,\th)+\frac{\stackrel{1}{\t}(u,\th)}{r}
        +\frac{\stackrel{2}{\t}(u,\th)}{r^2}+{\cal O}(r^{-3})   \ ,
\label{TrBondiSchw}\\
\ffi  &=& %F_0
         \f +\stackrel{o}{f}(u,\th)+\frac{\stackrel{1}{f}(u,\th)}{r}
          +\frac{\stackrel{2}{f}(u,\th)}{r^2}
        +{\cal O}(r^{-3})   \ .\nn
\EEA

Comparing  the~resulting metric expansions 
 with the expansions (\ref{expanse}) we obtain differential equations for 
coefficients entering the~asymptotic transformation (\ref{TrBondiSchw}).
Since these equations are lenghty we present only their
solutions in App.~\ref{ap-bondi}. 
Their integrability condition (obtained comparing (\ref{t1}) and (\ref{t1u})) 
turns out to be the~same as in the~nonspinning case \cite{bicakBS}
\BDM
\b,_{\stackrel{o}{\p}} \stackrel{o}{\p} 
            +\b,_{\stackrel{o}{\t}}\tan\stackrel{o}{\t}=0\ ,\mm
\mbox{or equivalently}\mm \l_0,_U U+\l_0,_\Theta \tan\Theta=0\ ,
\EDM
where we used the~relations $\b,_u=\b,_{\stackrel{o}{\p}} 
\stackrel{o}{\p},_u$ and
$\b,_\th=\b,_{\stackrel{o}{\t}} \stackrel{o}{\t},_\th
     +\b,_{\stackrel{o}{\p}}\stackrel{o}{\p},_\th$.
Solving Eqs.~(\ref{foth}), (\ref{toth}), (\ref{pou}), and (\ref{tosin}) 
one may infer
the~first order coefficients in the~expansions (\ref{TrBondiSchw})
\BEA
\stackrel{o}{\t} &=&2\arctan \lvhz {\rm e}^{-\n}
                       (\tan \pul\th)^{K %F_0
                          }\pvhz\ ,\label{to}\\
q               &=& \frac{1}{\sqrt{K} %F_0
                               }\sin\pul\th\cos\pul\th 
                   \lvhz {\rm e}^\n 
             ({\rm cot} \pul\th)^{K %F_0
                            }
                 + {\rm e}^{-\n }
             (\tan \pul\th)^{K %F_0
                        }\pvhz\ ,\label{q}\\
\stackrel{o}{f}&=& \frac{a_0\a}{K}
            {\rm ln}\lvvkz\frac{\sin\stackrel{o}{\t} }
                {1+\cos\stackrel{o}{\t} }
        \pvvkz \ ,\label{fo}\\
\stackrel{o}{\p},_u&=&\frac{1}{\b q}\ ,\label{po}
\EEA
where $K\rov \frac{1+a_0^2\a^2}{\a}$ and $\n$ is an arbitrary constant.
The~axis (which is the~same in both coordinates, i.e. 
$\Theta=\ 0$, $\p$ $\msip$ $\th=\ 0$, $\p$) is singular for $K\not= 1$
as $q$ goes to $0$ for $K<1$ and to $\infty$ for $K>1$ there.
Since the~coordinate system $\{ t$, $\rho$, $z$, $\ffi\}$ 
can be chosen in such a way that $a_0=0$, we present here news functions 
for $a_0=0$ and the~case $a_0\not= 0$ is given in App.~\ref{ap-bondi}.
From Eqs.~(\ref{c}) and (\ref{d}) we obtain the~news functions 
\BEA
c,_u&=& \pul\m_1,_u-\frac{{q,_\th}^2}{2q^2}
  -\frac{q,_\th{\rm cot}\stackrel{o}{\t}}
                              {q^2\sqrt{\a}}
                 +\frac{1}{2q^2\b\sin^2\stackrel{o}{\t}}
                -\frac{1}{2}-\frac{ {\rm cot}^2\stackrel{o}{\t}}
                         {2q^2\a}\ ,\label{cu}\\
%----------------------------------------------------
d,_u&=&  -\pul \a a_1,_u\ .\ \label{du}
\EEA
Having the~news functions of the~system one can compute 
the~Bondi mass, see \cite{ajajibi}.

For a special case $K=\a=1$,  
i.e. for a regular axis, we get from  (\ref{to}), (\ref{q}), and (\ref{fo}) 
\BEA
\stackrel{o}{\t}&=&2\arctan \lvkz {\rm e}^{-\n}
                    \tan \pul\th\pvkz\ ,\nn\\
q               &=& \cosh\n+ \cs \sinh\n \ ,\\
\stackrel{o}{f}&=& 0\ .\nn
\EEA
Coordinate systems with different $\n$ are connected by Lorentz 
transformations along the~axis belonging to the~Bondi-Metzner-Sachs 
group and thus as in \cite{bicakBS} 
we may without loss of generality put $\n=1$ which 
implies $q=1$ and $\stackrel{o}{\t}=\th$. Then the~coefficient 
$\stackrel{o}{\p}$ can be computed from the~relation 
\BE
\int {\rm e}^{\lambda_0 (\stackrel{o}{\p},\ \th)}{\rm d}\stackrel{o}{\p}
=u+\o(\th)\ \label{pospec}
\EE
obtained from Eq.~(\ref{po}).
The~function $\o(\th)$ in (\ref{pospec})
represents a supertranslation also belonging  to  the~Bondi-Metzner-Sachs 
group and thus it may be again put equal to zero without loss of generality. 
Finally the~news functions (\ref{cu}) and (\ref{du}) read 
\BEA
c,_u&=&-\frac{1}{2\dsn}+\frac{1}{2\b\dsn}+\pul \m_1,_u=
\frac{1}{2\b\dsn}(1-\b+\m_1,_u \b\dsn )\ ,\label{speccu}\\
d,_u &=&-\pul a_1,_u\ .\label{specdu}
\EEA
For $a_1=0$ (\ref{speccu}) and (\ref{specdu})
reduce to news functions as given  
in \cite{bicakBS,BicTN} for the~nonrotating case.
%--------------------------------------------

\appendix
\section{Coordinate systems adapted to the~boost and rotation symmetries}
\label{ap-metric}

In the~nonradiative stationary region bellow the~roof, spinning brs 
metric can be transformed to the~stationary Weyl 
coordinates $\{ {\bar t}$, ${\bar \r}$, ${\bar z}$, ${\bar \ffi}\}$
with the~Killing vectors $\x=\der_{\bar \ffi}$, $\e=\der_{\bar t}$
and the~metric
\BE
{\rm d}s^2=-{\rm e}^{-2U}
        [ {\rm e}^{2\n} ( {\rm d} {\bar \r}^2+{\rm d}{\bar z}^2)
      +{\bar \r}^2{\rm d}{\bar \ffi}^2] 
+{\rm e}^{2U}({\rm d}{\bar t}+a{\rm d}{\bar \ffi})^2\ . \label{dsWeyl}
\EE
Vacuum Einstein's equations have the~form \cite{kramer}
\BEA
U,_{{\bar \r}{\bar \r}}+U,_{{\bar z}{\bar z}}+
\frac{U,_{\bar \r}}{{\bar \r}}
&=&-\frac{{\rm e}^{4U}}{2{\bar \r}^2} ({a,_{\bar \r}}^2+{a,_{\bar z}}^2)\ ,\nn\\
0&=&\lvkz {\rm e}^{4U}\frac{a,_{\bar \r}}{{\bar \r}}\pvkz,_{\bar \r}
   +\lvkz {\rm e}^{4U}\frac{a,_{\bar  z}}{{\bar \r}}\pvkz,_{\bar  z}\ ,\nn\\
\frac{\n,_{\bar \r}}{{\bar \r}} &=& {U,_{\bar \r}}^2-{U,_{\bar z}}^2
           -\frac{{\rm e}^{4U}}{4{\bar \r}^2}
          ({a,_{\bar \r}}^2-{a,_{\bar z}}^2)\ ,\label{ErceWeyl}\\
\frac{\n,_{\bar  z}}{{\bar \r}} &=& 2{U,_{\bar \r}}{U,_{\bar z}}
           -\frac{{\rm e}^{4U}}{2{\bar \r}^2}
               {a,_{\bar \r}}{a,_{\bar z}}\ .\nn
\EEA

%--------------------------------

Another appropriate coordinate system in the~stationary region
bellow the~roof is $\{ \g,\ \r,\ \b,\ \ffi\}$ with Killing vectors
$\x=\der_\ffi$, $\e=\der_{\g}$ and the~metric
%-----------------------------------
\BE
{\rm d}s^2=-{\rm e}^\l ({\rm d}\r^2 +{\rm d}\b^2)
-\r^2{\rm e}^{-\m}{\rm d}\ffi^2 
+\b^2{\rm e}^{\m}({\rm d}\g+a{\rm d}\ffi)^2\ ,\label{dsbetagama}
\EE
connected with the~stationary Weyl coordinates by 
(see (5.4), (5.6) in \cite{BicSchPRD})
\BDM
{\bar t}=\g\ ,\mm {\bar \r}=\r\b\ ,\mm {\bar z}-{\bar z}_0
=%\pm 
\frac{\b^2-\r^2}{2}\ ,\mm
{\bar \ffi}=\ffi\ ,\mm {\bar z}_0=\mbox{const} \ ,
\mm {\rm e}^{2U} =\b^2 {\rm e}^{\m}\ ,\mm
 {\rm e}^{2\n}=\frac{\b^2}{\r^2+\b^2}{\rm e}^{\m+\l}\ .     \nn
\EDM
%---------------------------------
Vacuum Einstein's equations 
read
\BEA
\m,_{\r\r}+\m,_{\b\b}+\frac{\m,_\r}{\r}+\frac{\m,_\b}{\b}
&=&-\frac{ \b^2}{\r^2} {\rm e}^{2\m}  ({a,_{\r}}^2+{a,_{\b}}^2)\ ,\nn\\
0&=&\lvkz \frac{\b^3}{\r}{\rm e}^{2\m}a,_{\r}\pvkz,_{\r}
   +\lvkz \frac{\b^3}{\r}{\rm e}^{2\m}a,_{\b}\pvkz,_{\b}\ ,\nn\\
(\r^2+\b^2) \l,_\r &=& (\r^2-\b^2)\m,_\r-2\r\b\m,_\b
-\pul\r\b^2({\m,_\b}^2-{\m,_\r}^2)+\r^2\b\m,_\r\m,_\b
\label{Ercebetagama}\\&&    
           +\frac{\b^4}{2\r}{\rm e}^{2\m}
   \lvkz {a,_\b}^2-{a,_\r}^2-2\frac{\r}{\b} a,_\r a,_\b\pvkz\ ,\nn\\
(\r^2+\b^2) \l,_\b &=& (\r^2-\b^2)\m,_\b+2\r\b\m,_\r
+\pul\r^2\b({\m,_\b}^2-{\m,_\r}^2)+\r\b^2\m,_\r\m,_\b
\nn\\ &&           
           -\pul \b^3{\rm e}^{2\m}\lvkz {a,_\b}^2-{a,_\r}^2
        +2\frac{\b}{\r} a,_\r a,_\b\pvkz\ .\nn
\EEA
The~stationary region of a brs spacetime, under the~roof, is composed
of two identical regions ($z>0$, $z>|t|$ and $z<0$, $z<-|t|$)
and each of them can be transformed to coordinates 
$\{ {\bar t}$, ${\bar \r}$, ${\bar z}$, ${\bar \ffi}\}$
or $\{ \g,\ \r,\ \b,\ \ffi\}$.
%-----------------------------------------

By further transformation (3.5) in \cite{BicSchPRD} 
to coordinates $\{ t$, $\r$, $z$, $\ffi\}$
\BDM
\tanh\g =\pm \frac{t}{z}\ ,\mm \b=\sqrt{z^2-t^2}\ ,\mm 
B\rov \b^2\ ,\mm A\rov \r^2\ ,\nn
\EDM
we arrive at the~metric (\ref{dsbrs}) where nonstationary region
above the~roof (again composed of two identical regions) 
appears as in the~nonspinning case \cite{BicSchPRD}.
%------------------------------------------

For examining regularity of the~axis it is convenient to transform
(\ref{dsbrs}) to coordinates $\{ t$, $x$, $y$, $z\}$, 
where $x=\r\cos\ffi$, $y=\r\sin\ffi$:
\BEA
{\rm d}s^2&=&-\frac{1}{x^2+y^2}\lvhz
            ({\rm e}^\l x^2+{\rm e}^{-\m}y^2){\rm d}x^2
            + ({\rm e}^\l y^2+{\rm e}^{-\m}x^2){\rm d}y^2
              +2xy({\rm e}^\l-{\rm e}^{-\m}){\rm d}x{\rm d}y\nn\\
&&     -\frac{z^2-t^2}{x^2+y^2}a^2{\rm e}^\m (-y{\rm d}x+x{\rm d}y)^2
              -2a{\rm e}^\m(-yz{\rm d}x{\rm d}t +yt{\rm d}x{\rm d}z 
           +xz{\rm d}y{\rm d}t -xt{\rm d}y{\rm d}z)\pvhz\label{dsxy}\\
&&
-\frac{1}{z^2-t^2}[( {\rm e}^\l z^2-{\rm e}^\m t^2) {\rm d}z^2
                  -2zt ({\rm e}^\l-{\rm e}^\m) {\rm d}z{\rm d}t
                  -({\rm e}^\m z^2-{\rm e}^\l t^2){\rm d} t^2]
\ .\nn
\EEA

%-------------------------------------------
Let us finally write down vacuum Einstein's equations for the~metric (\ref{dsbchi})
with the~Killing vectors $\x=\der_\ffi$, $\e=\der_\chi$
\BEA
\m,_{\r\r}-\m,_{bb}+\frac{\m,_\r}{\r}-\frac{\m,_b}{b}
&=&\frac{ b^2}{\r^2} {\rm e}^{2\m}  ({a,_{\r}}^2-{a,_{b}}^2)\ ,\nn\\
0&=&\lvkz \frac{b^3}{\r}{\rm e}^{2\m}a,_{\r}\pvkz,_{\r}
   -\lvkz \frac{b^3}{\r}{\rm e}^{2\m}a,_{b}\pvkz,_{b}\ ,\nn\\
(\r^2-b^2) \l,_\r &=& (\r^2+b^2)\m,_\r-2\r b\m,_b
 -\pul\r b^2({\m,_b}^2+{\m,_\r}^2)+\r^2 b\m,_\r\m,_b
           \label{Ercebchi}\\&&     
           +\frac{b^4}{2\r}{\rm e}^{2\m}
     \lvkz -{a,_b}^2-{a,_\r}^2+2\frac{\r}{b} a,_\r a,_b\pvkz\ ,\nn\\
(\r^2-b^2) \l,_b &=&- (\r^2+b^2)\m,_b+2\r b\m,_\r
-\pul\r^2 b({\m,_b}^2+{\m,_\r}^2)+\r b^2\m,_\r\m,_b
            \nn\\&&    
           +\frac{b^3}{2}{\rm e}^{2\m}
     \lvkz -{a,_b}^2-{a,_\r}^2+2\frac{b}{\r} a,_\r a,_b\pvkz \ .\nn
\EEA
The~coordinates $\{ b$, $\r$, $\chi$, $\ffi\}$ 
for the~nonstationary region above the~roof are analogical to coordinates
$\{ \g,\ \r,\ \b,\ \ffi\}$ (\ref{dsbetagama}) in the~stationary region
bellow the~roof.

As for (\ref{Ercemu})--(\ref{ErcelB}), 
in each set of Einstein's equations (\ref{ErceWeyl}), (\ref{Ercebetagama}),
and (\ref{Ercebchi}), the~first two are
integrability conditions for the~other two.

%-------------------------------------------

\section{Transformation of the~spinning brs metric to the~Bondi-Sachs coordinates}
\label{ap-bondi}

The~spinning brs metric (\ref{dsbrs})  with expansions
(\ref{rozvlma})--(\ref{betaalfa})  being transformed
to the~Bondi-Sachs coordinates with the~metric (\ref{ds})
using transformations (\ref{trURTh}), (\ref{TrBondiSchw}) and
compared with (\ref{expanse}) leads to lenghty
equations for coefficients of the~asymptotic transformation 
(\ref{TrBondiSchw}) and metric functions from (\ref{expanse}).
We present here only their solutions:
\BEA
(g_{u\f},\ r^2)&=&\ \ 0\mm\msip\ 
             \stackrel{o}{f},_u=\stackrel{o}{\t},_u
        \frac{a_0\a%^2
             }{K %(1+a_0^2\a^2)
         \sin\stackrel{o}{\t}}
                                  \ ,\nn\\
%--------------------------
(g_{uu},\ r^2)&=&\ \ 0\mm \msip\ \stackrel{o}{\t},_u=0=\stackrel{o}{f},_u\ ,\nn\\
(g_{uu},\ r^1)&=&\ \ 0\mm \msip\ q,_u=0\ ,\nn\\
%-------------------------
(g_{\th\f},\ r^2)&=&\ \  0\mm \msip\
                   \stackrel{o}{f},_\th=\stackrel{o}{\t},_\th
               \frac{a_0\a%^2
       }{K %(1+a_0^2\a^2)
          \sin\stackrel{o}{\t}}
                                  \ ,\label{foth}\\
%--------------------------
(g_{\th\th},\ r^2)&=&-1\mm \msip\  \stackrel{o}{\t},_\th=\pm\frac{
                \sqrt{K %\frac{1+a_0^2\a^2}{\a}
                     }}{q}
           \ (\mbox{we will use the~sign +}) \ ,\label{toth}\\
%--------------------------
(g_{ur},\ r^0)&=&\ \ 1\mm \msip\  \stackrel{o}{\p},_u=\frac{1}{\b q}
                \ ,\label{pou}\\
%---------------------------
(g_{r\f},\ r^0)&=&\ \ 0\mm \msip\
                       \stackrel{1}{f}=\frac{a_0\a}{q\sin^2\stackrel{o}{\t}}
                               \frac{\stackrel{1}{\t}q\sin\stackrel{o}{\t}
                             +\stackrel{o}{\p}\cos\stackrel{o}{\t}}
                                {K}\ ,\nn\\
%---------------------------
(g_{r\th},\ r^0)&=&\ \ 0\mm \msip\  \stackrel{1}{\t}=-
                    \frac{1}{\stackrel{o}{\t},_\th q\sin \stackrel{o}{\t}}
                   \lvhz \stackrel{o}{\p},_\th\b K\sin \stackrel{o}{\t}
                     +\stackrel{o}{\p}\stackrel{o}{\t},_\th
               \cos\stackrel{o}{\t} (1-\b K)
                                    \pvhz\ ,\label{t1}\\
%----------------------------
(g_{r\f},\ r^0)&=&\ \ 0\mm \msip\ \stackrel{1}{f}=\frac{a_0\a\b}
                           {q\stackrel{o}{\t},_\th\sin^2\stackrel{o}{\t}}
          (\stackrel{o}{\p}\stackrel{o}{\t},_\th\cos\stackrel{o}{\t}
              -\stackrel{o}{\p},_\th\sin\stackrel{o}{\t})\ ,\nn\\     
%----------------------------------------
(g_{u\th},\ r^1)&=&\ \ 0\mm \msip\  \stackrel{1}{\t},_u=
                             \frac{1}{\b q^3\stackrel{o}{\t},_\th 
              \sin \stackrel{o}{\t}}
                \lvhz q,_\th \b K\sin\stackrel{o}{\t} 
             +\stackrel{o}{\t},_\th q \cos\stackrel{o}{\t}
        (-1+\b K)\pvhz\ ,\label{t1u}\\
%--------------------------------------
(g_{\th\th}g_{\f\f},\ r^3)&=&\ \ 0\mm\msip\ \stackrel{o}{\s}=
                  \frac{1}{2{\stackrel{o}{\t},_\th}^3
            \sin^2\stackrel{o}{\t}}
      \lvsz {\stackrel{o}{\t},_\th}^3 
       (1-2\sin^2\stackrel{o}{\t})\stackrel{o}{\p}
             (1-\b K)
       +K\sin\stackrel{o}{\t} 
            \lvhz\b\stackrel{o}{\p},_\th 
        ({\stackrel{o}{\t},_\th}^2 \cos\stackrel{o}{\t}
               -\stackrel{o}{\t},_{\th\th} \sin\stackrel{o}{\t} )\nn\\
&&\mm\mm\mm\mm\mm\  +\stackrel{o}{\t},_\th 
         \lvkz\b,_\th (\stackrel{o}{\p},_\th \sin\stackrel{o}{\t} 
                    -\stackrel{o}{\p}\stackrel{o}{\t},_\th\cos\stackrel{o}{\t}   )
    +\b\stackrel{o}{\p},_{\th\th} \sin\stackrel{o}{\t} \pvkz  \pvhz
                \pvsz\ ,\nn\\
%--------------------------------------------------------
(g_{\f\f},\ r^2)&=&-\dsn\mm \msip\ \sin\stackrel{o}{\t}=
                 \pm\frac{\sn}{q 
                \sqrt{K}} 
\ ,\label{tosin}\\
%---------------------------------------------
(g_{\f\f},\  r^1)&=&2c\dsn\mm\msip\ c=-\frac{1}{2q\sin^2\stackrel{o}{\t} (1+a_0^2\a^2)}
           \lvhz 2\sin\stackrel{o}{\t} (1+a_0^2\a^2)
             (\stackrel{o}{\s}\sin\stackrel{o}{\t}
                     +\stackrel{1}{\t}q\cos\stackrel{o}{\t})\nn\\
&&\mm\mm\mm\mm\mm\mm\mm\mm  
         +\mu_1 q\sin^2\stackrel{o}{\t} (-1+a_0^2\a^2)
         +2a_0\a^2(a_0\stackrel{o}{\p}+a_1 q\sin^2\stackrel{o}{\t}  )\pvhz\nn\\
                    &&\ \mm\mm\mv \msip\  c,_u=
                -\frac{a_0\a }{K} a_1,_u
                +\frac{1-a_0^2\a^2}{2(1+a_0^2\a^2)}\m_1,_u
                 -\frac{{q,_\th}^2}{2q^2}\nn\\
&&\mm\mm\mm\mm\mm\mm\mm\mm\  
                 -\frac{q,_\th{\rm cot}\stackrel{o}{\t}\sqrt{K}}{q^2}
                 +\frac{1-a_0^2\a^2}{2q^2\b\sin^2\stackrel{o}{\t}
                      (1+a_0^2\a^2)}
                -\frac{1}{2}-\frac{K {\rm cot}^2\stackrel{o}{\t}}
                         {2q^2}\ ,\label{c}\\
%----------------------------------------------------
(g_{\th\f},\ r^1)&=&-2d\sn\mm\msip\ d=-\frac{1}{2qK\sin^2\stackrel{o}{\t}}
                         \lvhz (1-a_0^2\a^2)q\sin^2\stackrel{o}{\t}                        
         a_1    
                   +2qa_0 \m_1\sin^2\stackrel{o}{\t} 
                 +2a_0\stackrel{o}{\p}  \pvhz  \nn\\
         &&\ \mm\mm\mv \msip\  
       d,_u=-\frac{a_0}{q^2 K\b\sin^2\stackrel{o}{\t}}
                        -\frac{a_0}{K}\m_1,_u
                      -\frac{1-a_0^2\a^2}
               {2K}a_1,_u\ .\label{d}
%------------------------------------------------------
\EEA
Equations $(g_{uu},\ r^0)=1$, $(g_{u\th},\ r^2)=0$, $(g_{u\f},\ r^1)=0$,
$(g_{rr},\ r^{-1})=0$, $(g_{r\th},\ r^1)=0$,
$(g_{r\f},\ r^1)=0$, and $(g_{\th\th},\ r^1)=-2c$ are satisfied
identically.
%---------------------------------------

\begin{acknowledgments}
We are grateful to Professor J. {Bi\v c\' ak} for many inspiring
discussions on boost-rotation symmetric spacetimes. 
V.P. was supported by the~grant GACR-202/00/P030 and A.P.
by the~grant GACR-202/00/P031. The~work was also
partially supported by the~National Science Foundation
under the~grant NSF-Czech Rep. INT-9724783.
\end{acknowledgments}

%------------------------------

\end{document}